# scientific reports

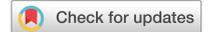

OPEN

# A case study of university student networks and the COVID-19 pandemic using a social network analysis approach in halls of residence

José Alberto Benítez-Andrades[1], Tania Fernández-Villa[2], Carmen Benavides[1]✉, Andrea Gayubo-Serrenes[3], Vicente Martín[2,4] & Pilar Marqués-Sánchez[5]

The COVID-19 pandemic has meant that young university students have had to adapt their learning and have a reduced relational context. Adversity contexts build models of human behaviour based on relationships. However, there is a lack of studies that analyse the behaviour of university students based on their social structure in the context of a pandemic. This information could be useful in making decisions on how to plan collective responses to adversities. The Social Network Analysis (SNA) method has been chosen to address this structural perspective. The aim of our research is to describe the structural behaviour of students in university residences during the COVID-19 pandemic with a more in-depth analysis of student leaders. A descriptive cross-sectional study was carried out at one Spanish Public University, León, from 23th October 2020 to 20th November 2020. The participation was of 93 students, from four halls of residence. The data were collected from a database created specifically at the university to "track" contacts in the COVID-19 pandemic, SiVeUle. We applied the SNA for the analysis of the data. The leadership on the university residence was measured using centrality measures. The top leaders were analyzed using the Egonetwork and an assessment of the key players. Students with higher social reputations experience higher levels of pandemic contagion in relation to COVID-19 infection. The results were statistically significant between the centrality in the network and the results of the COVID-19 infection. The most leading students showed a high degree of Betweenness, and three students had the key player structure in the network. Networking behaviour of university students in halls of residence could be related to contagion in the COVID-19 pandemic. This could be described on the basis of aspects of similarities between students, and even leaders connecting the cohabitation sub-networks. In this context, Social Network Analysis could be considered as a methodological approach for future network studies in health emergency contexts.

Adversities seem to have been a permanent reality in the last decade[1]. Their consequences cause damage to people's lives that deserve the attention of political leaders and researchers. In the context of any disaster, models of human behaviour are constructed that reflect the importance of relationships between actors, between actors and knowledge, and even between actors and beliefs[2].

The World Health Organization (WHO) declared the COVID-19 a global emergency on January 31, 2020[3]. It is one of the disasters that has had the greatest impact on our history. Recent studies have already shown that the COVID-19 pandemic appears to have an impact on mental health, leading to anxiety, depression, disturbed sleep quality and even increased perceptions of loneliness[4–11]. In the same sense, the impact of the pandemic has also "hit" young people, who go to school every day but who have seen their social relationships decline. The educational context was always present in the strategies implemented in previous pandemics. Some of the most common measures were the closure of schools to contain the transmission of influenza[12], support through informal networks on university campuses during the influenza A(H1N1) pandemic[13], and the need to increase knowledge on the pandemic, as it was found to influence everyday attitudes and practices[14].

[1]SALBIS Research Group, Department of Electric, Systems and Automatics Engineering, Universidad de León, Campus of Vegazana s/n, 24071 León, Spain. [2]The Research Group in Gen-Environment and Health Interactions (GIIGAS), Institute of Biomedicine (IBIOMED), Universidad de León, 24071 León, Spain. [3]Facultad de Ciencias de la Salud, Universidad de León, Campus of Vegazana s/n, 24071 León, Spain. [4]The Biomedical Research Centre Network for Epidemiology and Public Health (CIBERESP), 28029 Madrid, Spain. [5]SALBIS Research Group, Department of Nursing and Physiotherapy, Universidad de León, Campus de Ponferrada s/n, 24400 Ponferrada, Spain. ✉email: carmen.benavides@unileon.es

One of the measures that has had the greatest social impact in the COVID-19 pandemic has been the obligation to maintain a physical distance. Specifically, in the field of higher education, it seems to be remarkably complex and more difficult to carry out[15]. University campuses are of interest for studying social behaviour in the context of a pandemic. Numerous studies have shown how university students acquire healthy habits or, conversely, drug and alcohol consumption habits, depending on the type of relationships they have on campus and in the university residences[16,17].

However, there is a lack of studies that analyse the behaviour of university students based on their social structure during a pandemic. Therefore, a quantitative understanding of the behaviour of students in a health emergency situation is necessary as this information could be useful in making decisions about how to prepare for disasters. That is, how to act appropriately during and after an emergency of any kind, since interpersonal relationships, through which supportive and interdependent links are established and which are present in any emergency or disaster.

To address this structural perspective, the SNA method has been applied. The SNA is a distinctive perspective within the social and behavioural sciences. It is distinctive because it is based on the fact that relationships take place between interacting units[18]. For the SNA method, the unit of analysis is not the isolated individual, but the social entity made up of the actor with its possible connections, generating a structure[19]. The main perspective of the SNA focuses on the importance of the relationships between the units that interact in the social networks[18]. A social network is made up of a set of points or nodes that represent individuals or groups, and a set of lines that represent the interaction or otherwise, between the nodes, generating a social structure[20].

One of the most relevant premises of the SNA, for our study, is that it is not only assumed that individuals are connected through a structure, but that their goals and objectives are as well, because these are only achieved through connections and relationships[19,21,22]. Thus, the SNA could show us if university students with a more responsible goal form their own networks or mingle with their not-so-responsible peers. In relation to the groups, the actors influence and inform each other in a process that creates a growing homogeneity[21]. This perspective is of interest to this research.

The contacts between actors can be analyzed in two types of networks: sociocentric or complete networks and egocentric networks. The former includes an analysis between actors that belong to a delimited and previously defined census[23]. While the latter analyzes the structure that is generated between an ego and its contacts[24].

There is an extensive core of studies on SNA and health habits. Some of the most recent are related to contagion in substance use[25,26], physical activity[27], behavior related to the individual's low weight[28], engagement in university rooms[29] or eating behaviors[30] among others. SNA has even been applied to disaster scenarios such as droughts, floods, landslides, tsunamis, and cyclones[31]. No one thought that one year after this study, its results would be so useful for another scenario related to a major catastrophe such as the COVID-19 pandemic. Other recent studies shows a social network analysis approach in the problematic internet use among residential college students during COVID-19 lockdown[32] or associations between interpersonal relationships and mental health[33].

Based on the above, the purpose of this study was to analyse a community of university students and their structural behaviour in their university residences. Halls of residence form micro-communities where very close relationships develop, which can become a context of risk. In other words, university residences could become "places" that facilitate the spread of pandemics if adequate protocols are not followed. However, dormitories can also have a preventive value. Peer support behavioural patterns take place in them, among peers who are exposed to the same risks and circumstances. This sharing of similar situations can generate an enriching coping of personal experiences[34]. However, there is a lack of studies that analyse the structures of university students and their coping in crisis situations.

This study was conducted during one of the waves of the COVID-19 pandemic, where infection rates were at their highest. With the SNA methodology, the aim is to find answers to questions such as: What are the structural characteristics of the leading individuals in the dormitories? How are the contagion outcomes related to the structural positions in the network? For such questions, the proposed objectives were (i) to analyse the relationship between the students' network position and their outcomes with respect to the COVID-19 contagion, (ii) to describe the influential position of student leaders in the network, (iii) to analyse the Egonetwork of the most influential student leaders during the COVID-19 pandemic, and (iv) to visualise the relational behaviour of university students in the global network.

## Methods

**Study design.** A descriptive cross-sectional study was carried out at one Spanish Public University. The data was collected during one of the waves of the pandemic, specifically from 23th October 2020 to 20th November 2020.

The measures taken during the pandemic in the different regions of Spain were different, depending on the results of the contagion at each moment. At the time of carried out this study, teaching in the locality of the study was adapted to the situation. That is, there were limitations on the number of people, "mirror" classrooms, identification of QR, etc. In the town there was a limit to the number of people who could meet, pubs and discotheques had been closed, and there was a 10 pm curfew.

|  | Gender | | |
|---|---|---|---|
|  | Male N (%) | Female N (%) | Total (%) |
| Residence A | 18 (19.36) | 19 (20.43) | 37 (39.79) |
| Residence B | 8 (8.60) | 0 (0.00) | 8 (8.60) |
| Residence C | 21 (22.58) | 0 (0.00) | 21 (22.58) |
| Residence D | 16 (17.20) | 11 (11.83) | 27 (29.03) |
| Total (%) | 63 (67.74) | 30 (32.26) | 93 (100.0) |

**Table 1.** Sample characteristics.

**Setting and sample.** The participation was of 93 students, from 4 university residences. The characteristics of the sample can be seen in Table 1. Of the total participants, 32.26% were women and 67.74% men.

**Ethical consideration.** All participants received an informed consent form to participate in the study. Lastly, participants were offered the possibility of retracting consent once they had signed the form, without needing to provide a reason, and an email contact address was given should they require any further information. Participation was voluntary, and subject availability was respected at all times. All the participants that were involved in the study have given their informed consent to participate in this study.

The data for this study are considered health-related data. They comply with Directive 03/2020 of the European Data Protection Committee[35]. The researchers requested anonymised data from the responsible body of the university in charge of contacts COVID-19.

The study was approved by the Ethics Committee of the University of León (ETICA-ULE-008-2021).

**Data collection.** We collected the data from the database created at the university, SIVeULE, created for the follow-up of cases of COVID-19. This database collates the characteristics of the actors and their RT-qPCR result.

In the university there was a protocol to indicate norms and rules of (i) hygiene and preventive measures, (ii) what to do if you had symptoms, (iii) definitions of what was considered "close contact", "confinement", and " positive result ". There was support staff to collect data, deal with doubts, and assist both positive actors and confined actors. These people were called "trackers." The name defined their role because they identified the student's contacts that were positive, had symptoms, or had been "in close contact" with a positive person.

In the database, other data such as name, residence, gender, grade, name of contacts, and date and result of Polymerase Chain Reaction (PCR) test are also collected.

For the present study, the names were anonymized and registered in matrices for subsequent analysis using the SNA method.

The data obtained were used to construct a 93 × 93 matrix. The matrix was read as follows:

- For rows, "A nominates B";
- For columns, "A is nominated by B".

To carry out this study, the matrix has been symmetrized, determining that if A nominated B, B also nominated A. That is to say, it is an undirected matrix, since, if A had any contact with B, B also had contact with A.

**Data analysis.** For data analysis, we apply SNA to the 93 × 93 matrix. measures of centrality were applied to analyse *leadership* from a structural perspective. Centrality is a construct of the SNA that means the position in the network[18]. Previous researchers have applied SNA to the study of leadership, because they have conceptualized leadership as a process that starts from the collective and the interconnections[36–38]. For this study, the centrality measures selected were: degree, betweenness and eigenvector[18]:

The degree is the number of connections adjacent to an actor. Given the centrality of degree $d_i$ of the actor $i$ and $x_{ij}$ is the cell ($i, j$) of the adjacency matrix, then

$$d_i = \sum_j x_{ij}$$

Betweenness centrality is defined as the Extent to which an actor serves as a potential "go-between" for other pairs of actors in the network by occupying an intermediary position on the shortest paths connecting other actors. The formula for the centrality of node $j$ is given by the:

$$b_j = \sum_{i<k} \frac{g_{ijk}}{g_{ik}}$$

In this formula, $g_{ijk}$ represents the number of geodetic paths that connect $i$ and $k$ and through $k$ while $g_{ik}$ is the total number of geodetic paths between $i$ and $k$.

| Centrality | Residence | | | | F | p |
|---|---|---|---|---|---|---|
| | A<br>M ± SD | B<br>M ± SD | C<br>M ± SD | D<br>M ± SD | | |
| nDegree | 0.130 ± 0.079 | 0.112 ± 0.019 | 0.250 ± 0.000 | 0.132 ± 0.061 | 22.135 | < 0.001 |
| Eigenvector | 0.021 ± 0.046 | 0.003 ± 0.003 | 0.203 ± 0.000 | 0.013 ± 0.036 | 151.035 | < 0.001 |
| nBetweenness | 0.028 ± 0.059 | 0.011 ± 0.029 | 0.014 ± 0.000 | 0.026 ± 0.090 | 0.357 | 0.784 |

**Table 2.** One-way analysis of variance (ANOVA) of centrality measures (nDegree, eigenvector and nBetweenness) by residences.

Eigenvector centrality corresponds to the measure of actor centrality that takes into account the centrality of the actors to whom the focal actor is connected.

$$e_i = \lambda \sum_j x_{ij} e_j$$

Normalized measures were used.

The measures of centrality studied in the SNA have been the normalized degree (nDegree, the normalized degree centrality is the degree divided by the maximum possible degree expressed as a percentage), Eigenvector and nBetweenness (is the normalized betweenness centrality computed as the betweenness divided by the maximum possible betweenness).

To select the most leading students in the network, the measure of normalized *nBetweenness* was used[39]. This measure becomes more relevant during a pandemic, where the possibility of serving as a bridge or intermediary allows other networks to reach out, transferring good or bad practices and behaviors.

In order to have more information about the behaviour of the student leaders, the *Egonetwork* analysis of the most leading nodes for each component was carried out. Key players theory has been used to obtain this group of students displaying greater leadership[40]. Egonetwork studies the connections of a given node. This analysis in isolation is less comprehensive than the analysis of the entire network. But the researchers recommend this analysis combined with the analysis of the whole network to go deeper into the behaviour of certain nodes, depending on the objective of the research[24,34,41].

**Statistical analysis and visualisation.** IBM SPSS Statistics (26.0) software. was used for the statistical processing of the data. For the analysis of descriptive data, frequencies and percentages were used for the qualitative variables, whereas the mean and standard deviation were used for the quantitative variables. A chi-square test was carried out to verify whether there was a relationship between the groups, and the Student's t-test was used to compare the mean scores between the groups. An analysis of variance (ANOVA) was carried out to check the differences for continuous variables divided in groups. The UCINET tool, version 6.679[42] was used for the calculation of the SNA measurements. The tests carried out to study the normality of the distribution were Kolmogorov–Smirnov for populations of more than 55 individuals and the Shapiro–Wilk test for those less than or equal to 55. The level of statistical significance was set at 0.05. For qualitative analysis, a *visualization* of the global network will be carried out using Gephi, version 0.9.2, software. The key player tool has been used to calculate the key players of the network[43].

## Results

As shown in Table 2, there was a significant effect of residence on nDegree [F(3,89) = 22.135, p < 0.001] and Eigenvector [F(3,89) = 151.035, p < 0.001] and there was no significant effect of residence on nBetweenness [F = (3,89),p = 0.784].

Students in residence C have significantly higher degrees of centrality in nDegree and Eigenvector compared to the other residences. In the case of nBetweenness, students in residences A and D have higher values, although not significantly so.

Significant differences in all measures of centrality (nDegree, Eigenvector and nBetweenness) measures were found for the groups of people who tested positive for RT-qPCR (PCR+) versus those who tested negative for PCR (PCR-). The PCR + group of people had higher values of centrality than the PCR- group. The degrees of significance of these differences are shown in Table 3.

Significant differences were found between leaders and non-leaders calculated with the three measures of centrality and the prevalence of people who tested positive or negative for PCRs. Leaders had a higher percentage of people in the PCR + group compared to non-leaders. The degrees of significance of these differences are shown in Table 4.

Figure 1A shows the nodes of the study network highlighting in each colour which residence each one belongs to (A,B,C or D). In Fig. 1B the same network can be seen but the nodes with PCR + appear in red and and the nodes with PCR- in green. The distribution of the network allows us to appreciate the 4 different residences. The size of the nodes is represented by the nBetweenness of each node.

Figure 2 shows the network highlighting the trajectories of the three most important key players. The edges coming out of these key players are thicker than the others. Furthermore, the key players are numbered in order of importance in the network (1, 2 and 3). The size of the nodes is represented by the nBetweenness of each node.

|  | PCR+ M±SD | PCR− M±SD | Centrality | |
|---|---|---|---|---|
|  |  |  | t | p |
| nDegree | 0.340±0.136 | 0.152±0.093 | −7.828 | <0.001 |
| Eigenvector | 0.121±0.104 | 0.092±0.107 | −1.299 | <0.001 |
| nBetweenness | 0.026±0.059 | 0.004±0.017 | −2.545 | 0.013 |

**Table 3.** Independent-samples t-test of centrality measures (nDegree, eigenvector and nBetweenness) by PCR+ and PCR-.

|  | PCR+ N (%) | PCR− N (%) | Chi square tests of independence | |
|---|---|---|---|---|
|  |  |  | $\chi^2$ | p |
| **Leaders by degree** | | | | |
| Leaders | 10 (71.4) | 4 (28.6) | 5.249 | 0.222 |
| Non-leaders | 30 (38.5) | 48 (61.5) | | |
| **Leaders by eigenvector** | | | | |
| Leaders | 11 (78.6) | 3 (21.4) | 8.275 | 0.004 |
| Non-leaders | 29 (37.2) | 49 (62.8) | | |
| **Leaders by betweenness** | | | | |
| Leaders | 11 (78.6) | 3 (21.4) | 8.275 | 0.00 |
| Non-leaders | 29 (37.2) | 49 (62.8) | | |

**Table 4.** Results of the comparison between leaders and non-leaders who had PCR+ and PCR− applying the chi-square test of independence.

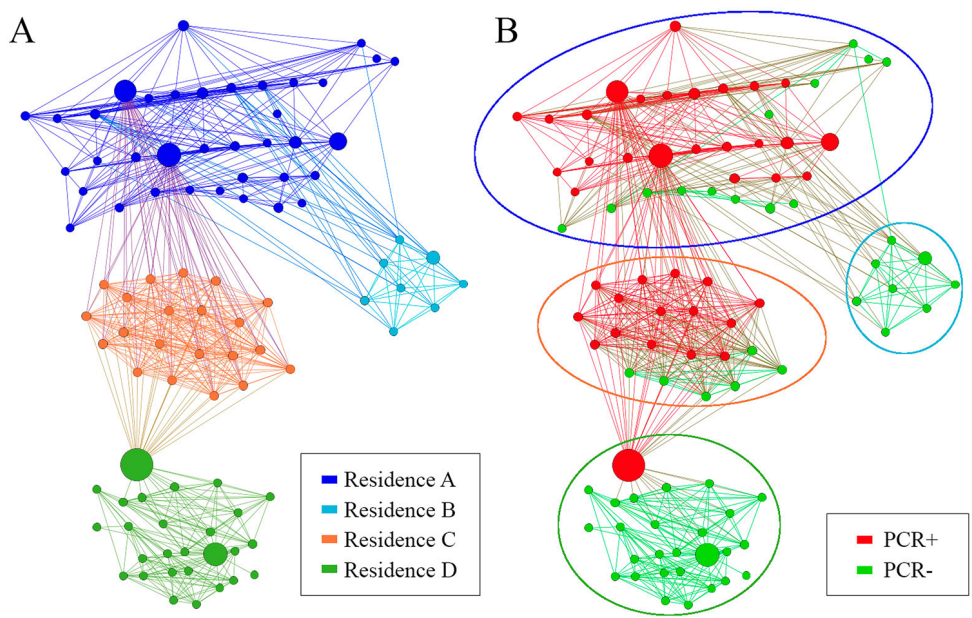

**Figure 1.** Graphs of the university student network differentiating a colour for each residence hall (**A**) and differentiating the positive and negative PCR groups (**B**).

Figure 3 shows the Egonetworks of the 3 key players in the network. Figure 3A shows the most important key player in the network. If this node were eliminated, the two components would be separated (those of the C and D residence). Figure 3B,C show the Egonetworks of the key players 2 and 3 respectively. These nodes are structurally very similar. If both nodes were removed from the network, there would no longer be a connection between residence C and residences A and B.

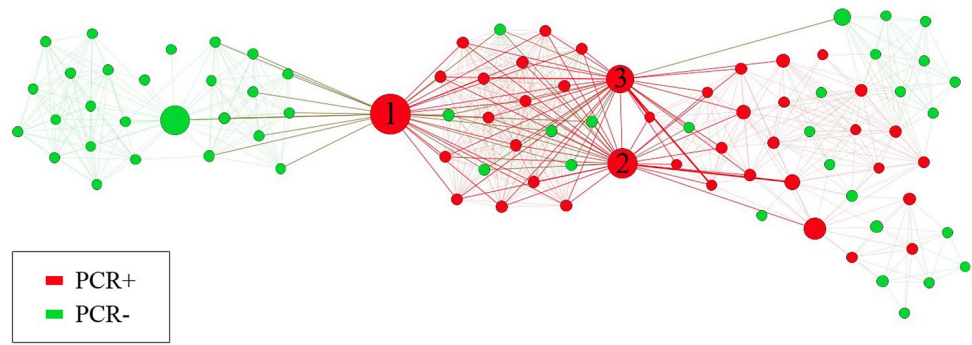

**Figure 2.** The network shown under the Atlas 2 distribution highlighting the 3 most important key players in the network.

## Discussion

This research contributes empirical evidence based on a social network approach to the development of the COVID-19 pandemic on university halls of residence. We have presented a study strategy and results, which link the relationship between the centrality of leaders and the outcome of pandemic infection. There is a significant core of research using the SNA methodology applied to the COVID-19 pandemic. However, there is a lack of research focusing on the structural responses of university students, a population of particular interest given their training experience. A university student "absorbs" experiences that are translated into behaviour, and transfers the resources obtained through their relationships.

Our results demonstrate the relationship between the centrality in the network of student leaders and the outcome of their infection (positive or negative). Not only could leaders spread pandemic behaviour towards their more local peers, they also seem to spread it to other halls of residence. This is demonstrated by the structure of betweenness. Leaders with a higher degree of betweenness could become key players, so that their presence or absence can disconnect the various components of the entire network. This could lead to a disconnection of the contagion process, both on a positive and negative level. The findings are the first to demonstrate that networks in university accommodation develop successful or unsuccessful responses to a pandemic. University managers should take these findings into account when developing response and behavioural strategies in pandemic or disaster situations. Strategies should be designed with a network rather than an individual approach.

Although our study did not ask about the relationship between the actors, we understand that the contacts established between the students are relationships of friendship or good classmates. We only analysed whether or not people had been in contact, during a state of lockdown. But obviously, with the SNA, we can visualise relational behaviours that would be more difficult to appreciate using other methodologies.

Our results show that student leaders have a high degree of centrality not only at the local level, i.e. in the component related to their accommodation, but also at the level of the global network. Our results are in line with studies of Mehra et al.[36], who highlighted that the integration of a leader into the friendship network in one social circle can be related to the reputation of the leader in other social circles.

Leadership or reputation at the local level is related to the performance of the team, and leadership outside the team is what allows new opportunities to arise and new information to be disseminated[36]. In the case of university students in their accommodation, the aim is to have a friendly atmosphere and to collaborate in difficult moments, to motivate each other, etc. Our results shown a statistically significant relationship between leadership and the positive results of the COVID-19 tests. In this sense, previous studies have already found that having too many resources related to social capital in a group (such as centrality) could negatively affect the efficiency of the group[44]. In other words, the leader will exert an influence on his or her colleagues and this influence could "infect" a certain behaviour, in this case of responsibility or not in a state of health emergency.

Another aspect demonstrated in our research is that there is a similarity between student groupings in terms of their COVID-19 test results. That is, we observe groups where the results are all positive (nodes in red), and others where the results are negative (nodes in green). This finding, could be related to numerous previous studies where actors occupy similar social positions in the classroom. For example, the studies from[45] showed that stuttering students had the same social position as the rest of their peers, because both (stutterers and non-stutterers) tended to design their groups structurally the same.

Homophily theory indicates that individuals associate with those with whom they share aspects of similarity, such as similar beliefs, characteristics and behaviours, which occurs especially in young people and adolescents[46]. Therefore, this may partly justify why negative-test college students are more cohesive, and positive-test college students as well.

One of the measures implemented with the greatest impact in this COVID-19 pandemic has been social distancing or isolation. The closure of premises or the reduction in hours of places of leisure has led to this social, or rather physical, distancing, as it is physical contact that is avoided. Studies have shown that the reduction in contacts based on social networks that coexist in social bubbles, and the similarity between contacts, increase social distancing from other actors, and therefore decrease the risks of contagion[47]. But in the case of this research, university accommodation could not be considered as a bubble. We could think of them as big bubbles, where behavioural patterns become contagious, be they positive and negative ones. Therefore, in this sense, the directors of the centres should take note and plan different strategies according to the behaviour of the subnetworks. That is to say, promote those behaviours with negative results of contagion and intervene in those subnetworks with positive results. For this, and as explained previously, the best option would be to plan together with the leaders.

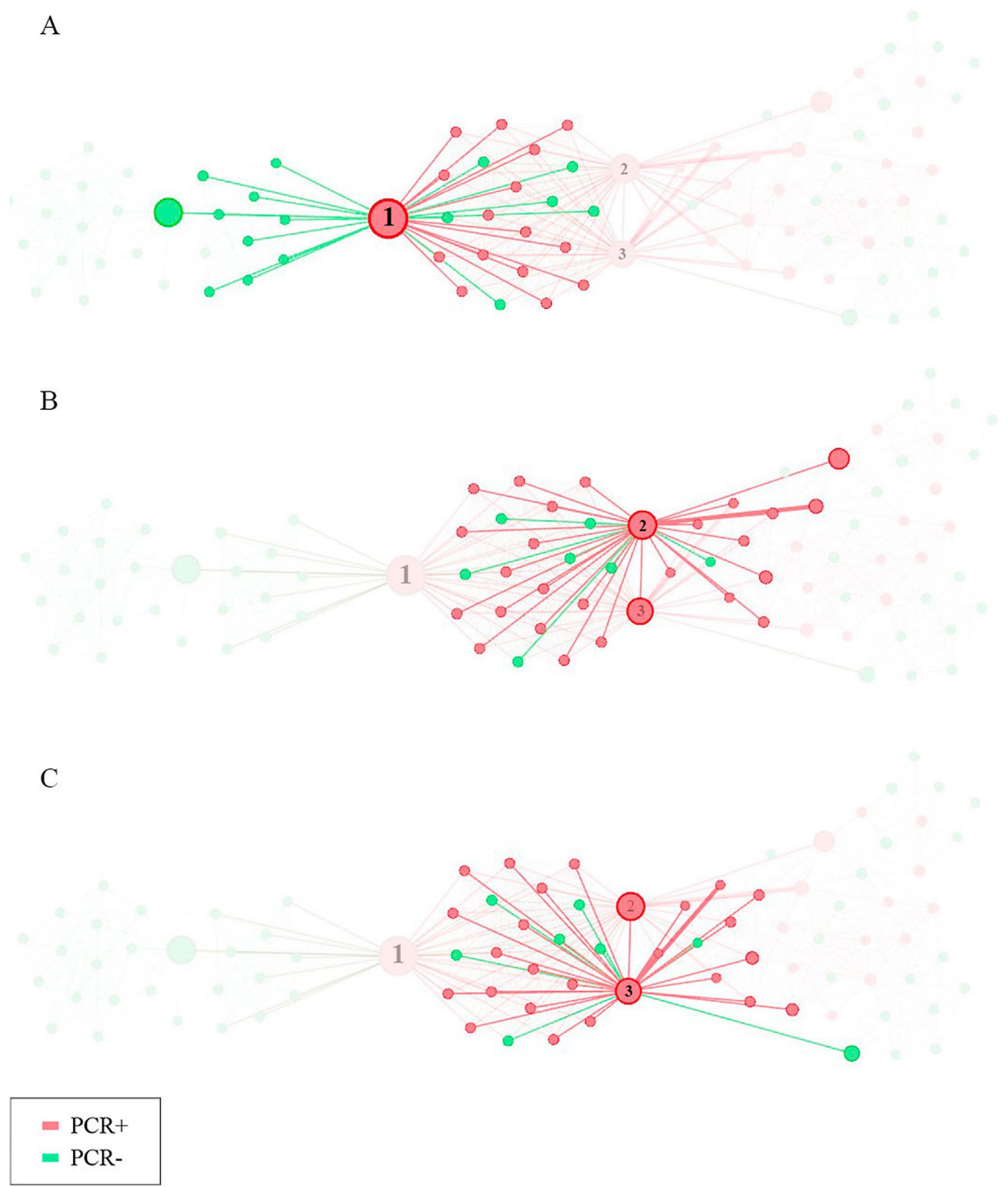

**Figure 3.** Egonetworks of the 3 main key players of the network.

Our results have shown that students with a high degree of *Betweenness* have a position in the network that gives them great leadership. In this sense, previous studies have used this structural metric as a predictor of leadership due to the strategic position that the actors have in the network and their role in bridging different networks[39,48].

For a better understanding of the role of these actors, in this research we analyse these university students on the basis of two more structural issues. On the one hand, which of them could have a key player role. Secondly, to analyse the Egonetwork of those students with a greater degree of centrality in each of the components.

As regards *key players,* our results showed that 3 students with a high degree of betweenness, i.e. with an intermediary role, had a key player structure. The importance of the key actor has been explained perfectly by Borgatti (2006)[40], describing both the negative and positive aspects. The negative is that the network, or networks, actually depend on these nodes, and cohesion between the networks would be diminished if these actors were to disappear[40]. This problem is greater when, in a public health context, we select a small number of individuals to contain a pandemic or to reduce the risk of contagion that links different networks. If these actors disappeared, the number of those infected would increase. As regards the positive role of these actors, they are ideal for spreading attitudes and behaviour, because they quickly gain access to different networks. Borgatti (2006) explains the importance of the structure of the key players, with the same relevance in very different contexts, such as terrorist networks or pandemic contexts[40]. In our case, our results are supported by the justification of this great researcher.

Our findings have shown that student leaders with a higher degree of Betweenness had a higher density than their peers in their *Egonetworks*. This could facilitate the transmission of social capital in a context such as the COVID-19 pandemic. These students, who serve as bridges, could become key actors with the ability to mobilize and coordinate social activity[49]. Their role is key for other colleagues, since they could serve as a "mirror" to "invite" appropriate behaviors in a health emergency. The key question that remains is, what behavior do they have? Structurally, the present investigation has demonstrated and justified that its position in the network is a model that could be disseminated among the rest of the actors.

To summarize the above, those responsible for universities must take into account the collective behavior of its networks. In a context such as the COVID-19 pandemic, the diffusion of behaviors is very relevant. Authors call for "urban intelligence" as a possible strategy to deal effectively with a pandemic. They understand that the impact of a health emergency is more than just a public health problem since it involves social risks and instability. This situation would be better dealt with by having the best that the social and community structure can offer, the so-called "urban intelligence"[50].

SNA could provide a set of terms and concepts to explain and describe social phenomena[51]. The method offers a distinctive approach to analysing leadership in disaster processes. Leaders could be like "builders" of social responses and the managers of the universities should take it into account for the intervention processes.

The most important limitations of this study should be considered for future research. For example, it would be of interest to carry out other analyzes focused more on the cohesion of the network and the behavior of the subgroups, in order to draw structural conclusions at the micro level. Future lines of research could focus on comparing the students' leadership in terms of structure with leadership as perceived by both them and their own peers.

## Conclusions

The present research has carried out a study with students in university residences. The aim has been to describe the structural behaviour of students in university residences during the COVID-19 pandemic, with a more in-depth analysis of student leaders. The specific objectives proposed to develop the research were to: (i) analyse the relationship between the position of students in the network and their results with respect to COVID-19 infection, (ii) describe the position of influence of student leaders in the network, (iii) analyzing the Egonetwork of the most influential student leaders on the COVID-19 pandemic, and (iv) visualise the relational behaviour of university students in the global network.

The main conclusions derived from the results are detailed below:

- The most central students in the network, had more positive results regarding COVID-19 infection.
- The leadership of the confined students was related to higher degree, eigenvector and betweenness.
- A small core of leaders are key players, so their role conditions the connection or disconnection between different components of the global network.
- Students with a key player structure show a similar Egonetwork if they belong to the same residence.
- There is a student leader with the maximum key player power structure, causing a total disconnection between networks if he/she disappears from the global network.

The findings show that strategies to cope with a disaster or pandemic need to be addressed through a network approach. University managers will need to have a profound understanding of students' relational behaviour. Only then will the most restrictive measures be effective. Responsible or irresponsible behaviour is transferred through the connections between students, so Social Network Analysis should be considered as a method of analyzing the evolution of a pandemic at the societal level. Any crisis involves contacts, but in a pandemic, contacts can transfer infection. Also in a pandemic, contacts can transfer habits and behaviours "passed on" by leaders, so that they allow for more effective coping. All of this can be analysed using SNA. Our study provides findings with an innovative approach, achieved with SNA. Among the limitations of the study it should be noted that the sample is very small (n = 93). This means that we cannot state categorically the representativeness of the results presented. However, the results could be used for future research where it is useful to analyse health emergency contexts as a network rather than analysing individuals in isolation.

## References

1. UNISDR. *Global Assessment Report on Disaster Risk Reduction Making Development Sustainable: The Future of Disaster Risk Management.* (2015).

 2. Rodrigueza, R. C. & Estuar, M. R. J. E. Social network analysis of a disaster behavior network: An agent-based modeling approach. in *Proceedings of the 2018 IEEE/ACM International Conference on Advance Social Networks Analysis Mining, ASONAM 2018* 1100–1107. https://doi.org/10.1109/ASONAM.2018.8508651 (2018).
 3. World Health Organization (WHO). Coronavirus disease 2019 (2019-nCOV) situation report-11. *WHO Bull.* 1–7 (2020).
 4. Bauer, L. L. *et al.* Associations of exercise and social support with mental health during quarantine and social-distancing measures during the COVID-19 pandemic: A cross-sectional survey in Germany. *medRxiv.* https://doi.org/10.1101/2020.07.01.20144105 (2020).
 5. Grey, I. *et al.* The role of perceived social support on depression and sleep during the COVID-19 pandemic. *Psychiatry Res.* **293**, 113452 (2020).
 6. Rozanova, J. *et al.* Social support is key to retention in care during Covid-19 pandemic among older people with HIV and substance use disorders in Ukraine. *Subst. Use Misuse* **55**(11), 1902–1904 (2020).
 7. Arendt, F., Markiewitz, A., Mestas, M. & Scherr, S. COVID-19 pandemic, government responses, and public mental health: Investigating consequences through crisis hotline calls in two countries. *Soc. Sci. Med.* **265**, 113532 (2020).
 8. Chirico, F. The role of health surveillance for the SARS-CoV-2 risk assessment in the schools. *J. Occup. Environ. Med.* **63**, e255–e256 (2021).
 9. Chirico, F. & Ferrari, G. Role of the workplace in implementing mental health interventions for high-risk groups among the working-age population after the COVID-19 pandemic. *J. Health Soc. Sci.* **6**, 145–150 (2021).
 10. Chirico, F. *et al.* Prevalence of anxiety, depression, burnout syndrome, and mental health disorders among healthcare workers during the COVID-19 pandemic : A rapid umbrella review of systematic reviews. *J. Health Soc. Sci.* **6**, 209–220 (2021).
 11. Shala, M., Çollaku, P. J., Hoxha, F., Balaj, S. B. & Preteni, D. One year after the first cases of COVID-19: Factors influencing the anxiety among Kosovar university students. *J. Health Soc. Sci.* **6**, 241–254 (2021).
 12. Glass, L. M. & Glass, R. J. Social contact networks for the spread of pandemic influenza in children and teenagers. *BMC Public Health* **8**, 1–15 (2008).
 13. Wilson, S. L. & Huttlinger, K. Pandemic flu knowledge among dormitory housed university students: A need for informal social support and social networking strategies. *Rural Remote Health* **10**, 1526 (2010).
 14. Yap, J., Lee, V. J., Yau, T. Y., Ng, T. P. & Tor, P. C. Knowledge, attitudes and practices towards pandemic influenza among cases, close contacts, and healthcare workers in tropical Singapore: A cross-sectional survey. *BMC Public Health* **10**, 442 (2010).
 15. Sheehan, M. M., Pfoh, E., Speaker, S. & Rothberg, M. Changes in social behavior over time during the COVID-19 pandemic. *Cureus* **23**, 15–20 (2020).
 16. Dibello, A. M. *et al.* HHS Public Access. **66**, 187–193 (2019).
 17. Walsh, A., Taylor, C. & Brennick, D. Factors that influence campus dwelling university students' facility to practice healthy living guidelines. *Can. J. Nurs. Res.* **50**, 57–63 (2018).
 18. Wasserman, S. & Faust, K. *Social Network Analysis: Methods and Applications*. Structural Analysis in the Social Sciences. https://doi.org/10.1017/CBO9780511815478 (Cambridge University Press, 1994).
 19. Lozares, C. L. Teoría de redes sociales. *Pap. Rev. Sociol.* **48**, 103 (1996).
 20. Barnes, J. A. Class and committees in a Norwegian Island parish. *Hum. Relat.* **7**, 39–58 (1954).
 21. Borgatti, S. P. & Foster, P. C. The network paradigm in organizational research: A review and typology. *J. Manag.* **29**, 991–1013 (2003).
 22. Robins, G. *Doing Social Network Research : Network-Based Research Design for Social Scientists*. (2015).
 23. Hanneman, R. A. & Riddle, M. *Introduction to Social Network Methods*. (2005).
 24. Wölfer, R., Faber, N. S. & Hewstone, M. Social network analysis in the science of groups: Cross-sectional and longitudinal applications for studying intra- and intergroup behavior. *Gr. Dyn. Theory Res. Pract.* **19**, 45–61 (2015).
 25. Hunter, R. *et al.* Social network interventions for health behaviours and outcomes: A systematic review and meta-analysis. *PLoS Med.* **46**, 1–25 (2019).
 26. Henneberger, A. K., Mushonga, D. R. & Preston, A. M. Peer influence and adolescent substance use: A systematic review of dynamic social network research. *Adolesc. Res. Rev.* https://doi.org/10.1007/s40894-019-00130-0 (2020).
 27. Prochnow, T., Delgado, H., Patterson, M. S. & Meyer, M. R. U. Social network analysis in child and adolescent physical activity research: A systematic literature review. *J. Phys. Act. Heal.* **17**, 250–260 (2020).
 28. Zhang, S., de la Haye, K., Ji, M. & An, R. Applications of social network analysis to obesity: A systematic review. *Obes. Rev.* **19**, 976–988 (2018).
 29. Fernández-Martínez, E. *et al.* Social networks, engagement and resilience in university students. *Int. J. Environ. Res. Public Health* **14**, 1488 (2017).
 30. De Rosis, S., Pennucci, F. & Seghieri, C. Segmenting adolescents around social influences on their eating behavior: Findings from Italy. *Soc. Mar. Q.* **25**, 256–274 (2019).
 31. Shehara, P. L. A. I., Siriwardana, C. S. A., Amaratunga, D. & Haigh, R. Application of social network analysis (SNA) to identify communication network associated with multi-hazard early warning (MHEW) in Sri Lanka. in *MERCon 2019—Proceedings, 5th International Multidisciplinary Moratuwa Engineering Research Conference* 141–146. https://doi.org/10.1109/MERCon.2019.8818902 (2019).
 32. Xia, Y., Fan, Y., Liu, T.-H. & Ma, Z. Problematic Internet use among residential college students during the COVID-19 lockdown: A social network analysis approach. *J. Behav. Addict.* https://doi.org/10.1556/2006.2021.00028 (2021).
 33. Zhang, S., Li, Y., Ren, S. & Liu, T. Associations between undergraduates' interpersonal relationships and mental health in perspective of social network analysis. *Curr. Psychol.* https://doi.org/10.1007/s12144-021-01629-3 (2021).
 34. Jariego, I. M., Ramos, D. H. & Lubbers, M. J. Efectos de la estructura de las redes personales en la red sociocéntrica de una cohorte de estudiantes en transición de la enseñanza secundaria a la universidad. *Univ. Psychol.* **17**, 1–12 (2018).
 35. European Data Protection Board. *Guidelines 03/2020 on the Processing of Data Concerning Health for the Purpose of Scientific Research in the Context of the COVID-19 Outbreak*. Vol 13 (2020).
 36. Mehra, A., Smith, B. R., Dixon, A. L. & Robertson, B. Distributed leadership in teams: The network of leadership perceptions and team performance. *Leadersh. Q.* **17**, 232–245 (2006).
 37. Emery, C., Calvard, T. S. & Pierce, M. E. Leadership as an emergent group process: A social network study of personality and leadership. *Gr. Process. Intergr. Relat.* **16**, 28–45 (2013).
 38. Knaub, A. V., Henderson, C. & Fisher, K. Q. Finding the leaders: An examination of social network analysis and leadership identification in STEM education change. *Int. J. STEM Educ.* **5**, 26 (2018).
 39. De Brún, A. & McAuliffe, E. Social network analysis as a methodological approach to explore health systems: A case study exploring support among senior managers/executives in a hospital network. *Int. J. Environ. Res. Public Health* **15**, 511 (2018).
 40. Borgatti, S. P. Identifying sets of key players in a social network. *Comput. Math. Organ. Theory* **12**, 21–34 (2006).
 41. Borgatti, S. & Halgin, D. On network theory. *SSRN Electron. J.* https://doi.org/10.2139/ssrn.2260993 (2011).
 42. Borgatti, S. P., Everett, M. G. & Freeman, L. C. Ucinet for windows: Software for social network analysis. *Harvard Anal. Technol.* https://doi.org/10.1111/j.1439-0310.2009.01613.x (2002).
 43. Borgatti, S. *KeyPlayer*. (2003).
 44. Oh, H., Chung, M.-H. & Labianca, G. Group social capital and group effectiveness: The role of informal socializing ties. *Acad. Manag. J.* **47**, 860–875 (2004).

45. Adriaensens, S., Van Waes, S. & Struyf, E. Comparing acceptance and rejection in the classroom interaction of students who stutter and their peers: A social network analysis. *J. Fluency Disord.* **52**, 13–24 (2017).
46. McMillan, C., Felmlee, D. & Osgood, D. W. Peer influence, friend selection, and gender: How network processes shape adolescent smoking, drinking, and delinquency. *Soc. Netw.* **55**, 86–96 (2018).
47. Block, P. *et al.* Social network-based distancing strategies to flatten the COVID-19 curve in a post-lockdown world. *Nat. Hum. Behav.* **4**, 588–596 (2020).
48. Valente, T. W. Social networks and health. *Vasa* https://doi.org/10.1093/acprof:oso/9780195301014.001.0001 (2010).
49. Meltzer, D. *et al.* Exploring the use of social network methods in designing healthcare quality improvement teams. *Soc. Sci. Med.* **71**, 1119–1130 (2010).
50. Lai, Y., Yeung, W. & Celi, L. A. Urban intelligence for pandemic response: Viewpoint. *JMIR Public Heal. Surveill* **6**, e18873 (2020).
51. Park, M., Lawlor, M. C., Solomon, O. & Valente, T. W. Understanding connectivity: The parallax and disruptive-productive effects of mixed methods social network analysis in occupational science. *J. Occup. Sci.* **28**, 287–307. https://doi.org/10.1080/14427591.2020.1812106 (2020).

### Author contributions
V.M. and T.F.-V. conceived the project. J.A.B.-A., P.M.-S. and C.B. performed the analytical calculations. A.G.-S., and J.A.B.-A. performed all the numerical calculations. J.A.B.-A. and P.M.-S. wrote a first draft of the manuscript. All authors reviewed and edited the manuscript.

### Funding
This research received no external funding. This research received no specific grant from any funding agency in the public, commercial, or not-for-profit sectors.

### Competing interests
The authors declare no competing interests.

### Additional information
**Correspondence** and requests for materials should be addressed to C.B.

**Reprints and permissions information** is available at www.nature.com/reprints.

**Publisher's note**  Springer Nature remains neutral with regard to jurisdictional claims in published maps and institutional affiliations.